\documentclass{article}
\usepackage{ijcai24}

\usepackage{times}
\usepackage{soul}
\usepackage{url}
\usepackage[hidelinks]{hyperref}
\usepackage[utf8]{inputenc}
\usepackage[small]{caption}
\usepackage{graphicx}
\usepackage{amsmath}
\usepackage{amsthm}
\usepackage{booktabs}
\usepackage{algorithm}
\usepackage{algorithmic}
\usepackage[switch]{lineno}
\usepackage[table]{xcolor}
\usepackage{amssymb}
\usepackage{makecell}
\usepackage{diagbox}
\linenumbers

\title{DFB: A Data-Free, Low-Budget, and High-Efficacy Clean-Label Backdoor Attack}

\author{
	Binhao Ma$^1$
	\and
	Jiahui Wang$^1$\and
	Dejun Wang$^1$\And
	Bo Meng$^1$$^*$\\
	\affiliations
	$^1$South-Central Minzu University\\
	\emails
	1145183478@qq.com,
	1459006201@qq.com,
	380760195@qq.com,
	mengscuec@gmail.com$^*$
}

\begin{document}
	\nolinenumbers
	\maketitle
	
	\begin{abstract}
		In the domain of backdoor attacks, accurate labeling of injected data is essential for evading rudimentary detection mechanisms. This imperative has catalyzed the development of clean-label attacks, which are notably more elusive as they preserve the original labels of the injected data. Current clean-label attack methodologies primarily depend on extensive knowledge of the training dataset. However, practically, such comprehensive dataset access is often unattainable, given that training datasets are typically compiled from various independent sources. 
		
		Departing from conventional clean-label attack methodologies, our research introduces DFB, a data-free, low-budget, and high-efficacy clean-label backdoor Attack. DFB is unique in its independence from training data access, requiring solely the knowledge of a specific target class. Tested on CIFAR10, Tiny-ImageNet, and TSRD, DFB demonstrates remarkable efficacy with minimal poisoning rates of just 0.1\%, 0.025\%, and 0.4\%, respectively. These rates are significantly lower than those required by existing methods such as LC, HTBA, BadNets, and Blend, yet DFB achieves superior attack success rates. Furthermore, our findings reveal that DFB poses a formidable challenge to four established backdoor defense algorithms, indicating its potential as a robust tool in advanced clean-label attack strategies.
	\end{abstract}
	
	\section{Introduction}
	
	Over the past decades, deep learning has made significant advancements in various fields such as computer vision\cite{r1,r5}, natural language processing \cite{r2}, and gaming \cite{r3,r4}. However, state-of-the-art machine learning (ML) models require a substantial amount of data to achieve strong performance \cite{r6,r7}. Unfortunately, in the current landscape of large models and extensive datasets, neural networks have been demonstrated to be vulnerable to attacks \cite{r34,r35,r8}. Among them, backdoor attacks are the most covert and dangerous.
	
	The classical backdoor attack BadNets, Blend\cite{r8,r12} involves sticking a trigger to the source data and modifying its label to the target class. However, this type of attack is not very practical, as even basic data collectors could discern the correctness of labels.
	
	To improve the success rate of backdoor attacks through human inspection, recent research has introduced the concept of clean-label backdoor attacks. One proposed baseline is optimizing triggers across different training dataset classes. The Label Consistency (LC) attack \cite{r9} introduced two methods: one involves adding adversarial perturbations that make the original features more difficult to classify, and the other incorporates features from other non-target classes into target class samples through interpolation in the latent space of a Generative Adversarial Network (GAN). The Hidden Trigger Backdoor Attack (HTBA) \cite{r10} minimizes the distance in the feature space between perturbed inputs of the target class and inputs with triggers from non-target classes. However, they require access to the victim's training dataset for all classes. 
	
	Other methods, aimed at further reducing reliance on the training dataset, NARCISSUS \cite{r11} introduced an efficient clean-label attack employing a dataset from the target class. By training a classifier capable of accurately categorizing public out-of-distribution (POOD) samples, then fine-tunes on the target class in the training set. Although this approach significantly decreases the need for extensive access to the training dataset, enhances the success rate of attacks, and reduces the poisoning rate, it still necessitates access to the training data.
	
	Overall, the effectiveness of existing clean-label attacks largely relies on knowing training data from all classes or at least some portion of it. However, data originates from multiple independent sources in many real-world scenarios, making accessing the training data quite challenging. For instance, pre-trained models can be adapted to new tasks with only a small amount of training data. The typical approach for users is to download pre-trained deep models and acquire additional images from the internet to fine-tune the model for their specific task. In such applications, attackers cannot access or control users' training data. However, to the best of our knowledge, the existing clean-label backdoor attacks largely relies on an understanding of the entire training set or a portion of it.
	
	\textbf{To the best of our knowledge, we are the first to conduct clean-label attack without accessing the training set}. In this paper, we propose 'DFB', a clean-label attack that requires no access to any training dataset and relies solely on knowledge of the target class. Leveraging Public Out-of-Distribution (POOD) samples, we use a decoder to accurately classify the target class against other classes. Subsequently, we generate a trigger that contains no features from the target class using an encoder. We find that this trigger effectively leads the victim network to associate the trigger with the target class. The distinction between our approach and existing methods is illustrated in Table \ref{tb1}.
	
	Our contributions are as follows:
	
	\begin{itemize}
		\item We introduce a novel clean-label attack that requires only knowledge of the target class for attack and does not rely on any training data information, Additionally, we propose two methods for injecting triggers. The overall workflow is shown in Figure \ref{fig2}.
		
		\item We compare our attack against existing clean-label attacks on the CIFAR10, Tiny-ImageNet, and TSRD datasets. Even at low poisoning rates of 0.1\%, 0.025\%, and 0.4\%, respectively, DFB in the absence of training data achieves a significantly higher success rate than LC, HTBA, BadNets, and Blend, \textbf{even when LC and HTBA access to the training set}. The results as shown in Table \ref{tb4}.
		
		\item We demonstrate that four commonly implemented defense mechanisms are ineffective in adequately mitigating our attack.

		\item We have made the code publicly available, along with a curated POOD dataset designed for conducting attacks. This initiative aims to foster and propel further research into vulnerabilities of this nature in deep neural networks. The source code available at: https://github.com/Magic-Ma-tech/DFB.
	\end{itemize}

	\begin{table}
		\centering
		\caption{Summary of assumptions across typical attack, where $D$ represents requiring data from all classes, and ${D_t}$ represents needing data from the target class only.}
		\label{tb1}
		\resizebox{0.48\textwidth}{!}{
			\begin{tabular}{c|ccccc}
				\hline
				\hline
				& \textbf{\makecell[c]{Train from \\ scratch}}  & \textbf{ALL-to-one} & \textbf{Absence of $D$} & \textbf{Absence of ${D_t}$} & \textbf{\makecell[c]{Low poison\\ ratio}} \\  \hline

				\makecell[c]{BadNets \\ \cite{r8}} &  \cellcolor{blue!10} $\checkmark$    & \cellcolor{blue!10} $\checkmark$   & \cellcolor{blue!10} $\checkmark$    & \cellcolor{blue!10} $\checkmark$      & \cellcolor{pink!20} $\times$ \\ 
				
				\makecell[c]{Blend \\ \cite{r12}} &  \cellcolor{blue!10} $\checkmark$    & \cellcolor{blue!10} $\checkmark$   & \cellcolor{blue!10} $\checkmark$  & \cellcolor{blue!10} $\checkmark$  &\cellcolor{pink!20} $\times$  \\ 
				
				\makecell[c]{ LC \\ \cite{r9}}   &  \cellcolor{blue!10} $\checkmark$     & \cellcolor{blue!10} $\checkmark$   & \cellcolor{pink!20} $\times$    & \cellcolor{pink!20} $\times$          & \cellcolor{pink!20} $\times$ \\ 
				
				\makecell[c]{ HTBA \\ \cite{r10}} & \cellcolor{pink!20} $\times$      &\cellcolor{pink!20} $\times$   &  \cellcolor{pink!20} $\times$    & \cellcolor{pink!20} $\times$          & \cellcolor{pink!20} $\times$ \\ 
				
				\makecell[c]{NARCISSUS \\ \cite{r11}}  &  \cellcolor{blue!10} $\checkmark$    & \cellcolor{blue!10} $\checkmark$   & \cellcolor{blue!10} $\checkmark$    & \cellcolor{pink!20} $\times$          & \cellcolor{blue!10} $\checkmark$  \\ 
				
				\makecell[c]{\textbf{DFB} \\ \textbf{(Ours)} } & \cellcolor{blue!10} $\checkmark$     & \cellcolor{blue!10} $\checkmark$   & \cellcolor{blue!10} $\checkmark$    & \cellcolor{blue!10} $\checkmark$          & \cellcolor{blue!10} $\checkmark$ \\ \hline \hline
				
			\end{tabular}
		}

	\end{table}
	
	\begin{table*}[tb]
		
		\caption{Comparison of poison DFB's generation of two types of clean-label triggers with other methods (average). "Fixed" corresponds to the fixed\_trigger, while "Encoder" corresponds to the encoder\_trigger.}
		\label{tb4}
		\centering
		\resizebox{1\textwidth}{!}{
			\begin{tabular}{cccccccccc}
				
				\hline
				\hline
				Name & Clean & \makecell[c]{ HTBA \\ \cite{r10}} & \makecell[c]{BadNets-c \\ \cite{r8}}&\makecell[c]{BadNets-d \\ \cite{r8}} & \makecell[c]{Blend-c \\ \cite{r12}}& \makecell[c]{ Blend-d \\ \cite{r12}} &\makecell[c]{ LC\\  \cite{r9}}& \textbf{Ours(Fixed)} & \textbf{Ours(Encoder)}\\ \hline
				
				\makecell[c]{\textbf{Attack}\\ \textbf{Knowledge}} & \diagbox{}{}  & \cellcolor{pink!20} Training Data&\cellcolor{blue!10} \textbf{Absence of} $D$ &\cellcolor{blue!10} \textbf{Absence of} $D$&\cellcolor{blue!10} \textbf{Absence of} $D$ &\cellcolor{blue!10} \textbf{Absence of} $D$& \cellcolor{pink!20} Training Data  &\cellcolor{blue!10} \textbf{Absence of} $D$ &\cellcolor{blue!10} \textbf{Absence of} $D$ \\ \hline
				
				\multicolumn{10}{c}{ \textbf{\raisebox{-1ex}{CIFAR-10 results 0.1\% poison ratio (50 images)}}} \\ [2ex]
				
				\hline
				
				ACC & 95.54 & 95.29 & 95.33 &95.3& 95.4 &95.5& 95.4  & 95.52 & 95.53 \\ 
				Tar-ACC & 97.3 & 96.8 & 97.3 &97.1 &97.3 &97& 96.9  & 97.3 & 97.3 \\ 
				ASR&	0.43&	5.32&	0.42&93.7	&0.31&86.4&	16.3&	\textbf{98.8}&\textbf{97.6} \\ \hline
				
				\multicolumn{10}{c}{\textbf{\raisebox{-1ex}{Tiny-ImageNet results  0.025\% poison ratio (25 images)}}}\\ [2ex]
				
				\hline
				ACC&	69.23&	68.2&	68.73&	68.7&69.2&69&	69.13&	69.12&	68.9\\ 
				Tar-ACC&	70&	68&	70&	70&70&68	&68&	70&	68 \\ 
				ASR&	 0.18& 1.7	&  0.4&1.4&	 0.23&0.9&	 0.83& \textbf{77.8}&	\textbf{76.8}\\ \hline
				
				\multicolumn{10}{c}{\textbf{\raisebox{-1ex}{TSRD results 0.4\% poison ratio (20 images)}}}\\ [2ex]
				
				\hline
				
				ACC&	83.3&	82.2&	82.3&83	&83&82&	82&	83&	82.9\\ 
				Tar-ACC&	93&	90&	90&92	&92&90&	90&	92	&92\\ 
				ASR&	 0.2&	 2.6&	 1.9& 41.3&1.7&23.5& 2.4& \textbf{73.4} &\textbf{70.8}\\ \hline
				\hline

		\end{tabular}}
	\end{table*}

	\section{Related Work}
	\subsection{Backdoor Attacks}
	
	Dirty-label attacks have been extensively explored. For instance, \cite{r8} proposed using a pasted square as a backdoor trigger in images. \cite{r12} introduced a physical-world backdoor attack by incorporating objects into images as triggers. Subtle noise was utilized by \cite{r13} for covert backdoor attacks, while \cite{r11} developed smooth backdoor triggers to evade defenses. Additionally, references \cite{r14}, \cite{r15}, \cite{r16}, \cite{r36}, \cite{r17}, and \cite{r38} have contributed to enhancing the adaptability backdoor triggers. However, these approaches typically involve changing labels to the target class, making the poisoned inputs appear mislabeled to humans and easily detectable upon manual inspection.
	
	To improve the stealth of backdoor attacks, recent research has shifted towards clean-label backdoor attacks. We categorize these into two types: one requiring access to training sets for all classes, and the other requiring access to only a portion of the dataset.
	
	The Label Consistency (LC) attack \cite{r9} adds adversarial perturbations and interpolates in the latent space of a GAN. The Hidden Trigger Backdoor Attack (HTBA) \cite{r10} minimizes the distance between perturbed inputs of the target class and non-target class inputs with triggers in the feature space, using a pre-trained feature extractor. Both methods typically require access to the full training dataset of the victim's categories. In contrast, NARCISSUS \cite{r11} presents an efficient clean-label attack requiring only a target class dataset.
	
	However, these approaches still involve accessing a dataset, a challenge in real-life scenarios. To minimize the need for dataset access and reduce attack costs, we propose a novel method that does not require any training dataset access, relying solely on knowledge of the target label.

	\subsection{Backdoor Defense}
	
	A range of defense solutions against backdoor attacks has been developed in recent research. \cite{r18} proposed using pruning techniques to eliminate hidden backdoors in Deep Neural Networks (DNNs) by selectively removing specific neurons. \cite{r19} explored similar concepts, advocating for the removal of neurons with unusually high activation values, based on the $l\infty$ norm of activation mappings from the final convolutional layer. The pioneering trigger synthesis-based defense, Neural Cleanse, was introduced by \cite{r20}. This method involves deriving potential trigger patterns for each class, and then using an anomaly detector to determine the synthetic trigger pattern and its target label. Further research, as mentioned in \cite{r21}, \cite{r22}, and \cite{r23}, has focused on different approaches to generate potential triggers and conduct anomaly detection. \cite{r37} proposed active defenses against multi-domain backdoor attacks. The SentiNet method \cite{r24} detects triggers using spectral features, employing Grad-CAM \cite{r25} to identify crucial regions for each class. Additionally, STRIP \cite{r26} employs perturbations to detect potential backdoor triggers, using boundary analysis to determine trigger regions.
	
	In our experiments, we assess the effectiveness of our newly proposed 'DFB' attack against these four representative defense solutions.

	\section{DFB}
	
	\subsection{Threat Model}
	
	In our scenario, we consider a victim who trains a machine learning model (victim model) on a dataset aggregated from multiple sources. However, unlike \cite{r9,r10,r11}, the adversary is unable to access any of the victim's training datasets. They only possess knowledge about one specific category within the victim's training task and can supply data to the victim.
	
	\textbf{Attacker's Knowledge.} We assume that the attacker possesses solely knowledge about one category label within the victim's learning task. Taking a classification task as an example, we presume the attacker is aware that the victim is training a classifier capable of recognizing frogs. Consequently, the attacker has the opportunity to gather additional samples, either related or unrelated to the target category. It is important to note, however, that there is no guarantee these extra samples are drawn from the same distribution as the actual training data. For instance, if the attacker knows that the victim needs to train a classifier for classifying frogs, they might collect supplementary frog images from the internet or daily life. These images, however, could be self-captured or of an animated style, and might not align with what the victim uses. We refer to these additional examples related to the learning task as Public Out-of-Distribution (POOD) examples \cite{r11}. Given the prevalence of such data in many common learning tasks, particularly in image and language domains \cite{r27}, understanding the vulnerability of machine learning models under this knowledge is crucial. In this paper, we posit that the attacker can access some POOD examples, yet these examples are strictly segregated from the original training data.

	\textbf{Adversary's Objectives.} In order to elude potential human inspection, the adversary's goals are as follows: 1. The poisoned examples must have clean labels, indicating that the poisoned input and its corresponding label should appear coherent and consistent to a human observer. 2. The adversary's goal is to subtly modify the model $f_{\theta}$ so that it classifies any test input with a trigger $\delta $ into a designated target category T, without compromising its accuracy on clean inputs.
	
	\subsection{Problem Formalization}
	
	The effectiveness of existing clean label data stems from the intent to optimize triggers through datasets, thereby associating the trigger with the model for a successful attack. Compared to existing methods, we propose an indirect attack approach, using POOD to optimize triggers that can cover the features of the target category. We then inject this trigger into the target class. The neural network quickly establishes an association between this trigger and the target category. Therefore, we define the optimization of the trigger as follows:
	
	\begin{equation}
		\label{eq1} 
		{\delta ^*} = {f_\theta }((p{x^*} + \delta ),\overline {pt} )
	\end{equation}
	Where $p{x^*}$  represents the target class dataset in  ${D_{POOD}}$, and $\overline {pt} $ represents all labels except the target class label.
	
	\begin{figure*}[tb]
		\centering
		\includegraphics[width=1.9\columnwidth]{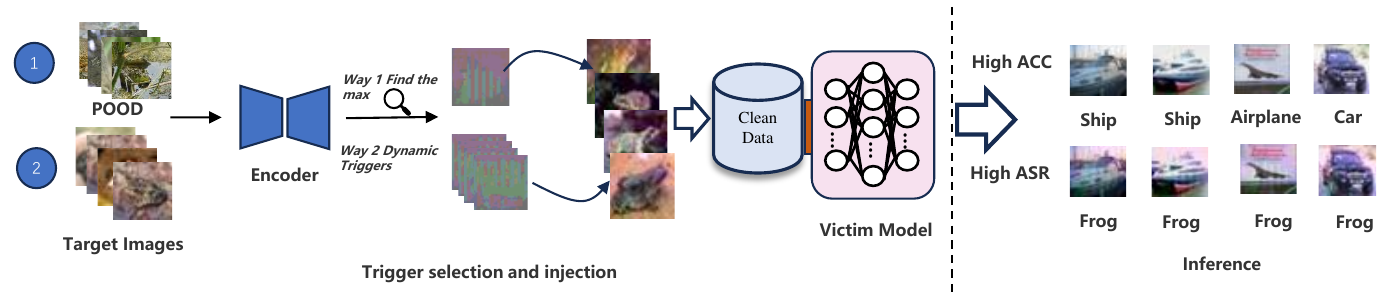}
		\caption{The DFB Pipeline employs two trigger selection methods: one to find loss-max triggers and the other to dynamically create and inject triggers into the clean dataset. In inference, it processes clean data normally but assigns the target label to trigger-embedded data.}
		\label{fig2}
	\end{figure*}

	\subsection{Attack Workﬂow}
	
	Next, we will provide a detailed explanation of the DFB attack method. The overall architecture of the DFB is depicted in Figure \ref{fig2}.
	
	\textbf{Step 1: Training the POOD target class decoder and encoder.} Initially, to meet the optimization goal in Formula \ref{eq1}, we collect the ${D_{POOD}}$ dataset. A challenge arises because ${D_{POOD}}$ contains multiple categories, making direct correlation with the formula's components difficult. Thus, our first step is to transform ${D_{POOD}}$ into a binary classification dataset, ${D_{POOD_Binary}}$, with labels comprising only the target class label $pt$ and a composite label $\overline {pt}$ for all non-target classes. This transformation enables the decoder to more effectively distinguish the target class features from others, unlike in a multi-class dataset. Therefore, we optimize the decoder by the following function:
	
	\begin{equation}
		\label{eq2}
		{f_{decode}} = \arg \min \sum {L({f_{decode}}((px),p{y_{true}}))} 
	\end{equation}
	Where $px$ represents the data in the training dataset ${D_{POOD\_Binary}}$, and ${py_{true}}$ includes instances where either $pt$ or $\overline {pt} $ belongs to ${py_{true}}$

	\textbf{Step 2: Training the Encoder for Trigger Generation Without Target Class Features.} 
	To optimize $\delta$, we employ the encoder. Target class data $p{x^*}$ is fed into the encoder to generate a corresponding residual $\delta$. This $\delta$ is then superimposed onto $p{x^*}$, creating $px_{\text{finish}}^*$. Feeding $px_{\text{finish}}^*$ into the decoder, we decode it to obtain its label. A loss function is constructed to refine $\delta$, tackling the optimization issue outlined in Formula \ref{eq1}. Post-optimization, the encoder is deemed capable of producing a residual ${\delta ^*}$ for images. When applied to the original image, this ${\delta ^*}$ can strip the target image of any features of the current class.
	
	
	\textbf{Step 3: Trigger Selection and Injection.} We propose two methods for trigger generation. The first method aims to identify triggers within the POOD dataset that maximize the likelihood of transforming the target image into a high-confidence non-target class using a specific loss function. The second method inputs $x_{\text{train}}$ from the training set into the encoder to generate the corresponding $\delta_{x_{\text{train}}}$. This $\delta$ is then injected into the training set. The algorithms for generating $\delta^*$ through these two trigger generation methods are detailed in Algorithm \ref{al2} and Algorithm \ref{al3}, respectively.

	\begin{algorithm}[tb]
		\caption{Fixed trigger generation}
		\label{al2}
		\textbf{Input}: ${f_{{\theta _{encoder}}}}$; ${f_{{\theta _{decoder}}}}$; ${D_{POOD\_Binary}}$; ${N_{p{x^*}}}$(The number of $p{x^*}$ in the POOD dataset )\\
		\textbf{Output}:${\delta^*}$(Fixed trigger)
		\begin{algorithmic}[1]
			\FOR {${N_{p{x^*}}}$}
			\STATE ${L_{\max }} = \max \left( L({f_{{\theta _{decoder}}}}(px_i^*), \overline {pt}) - \right.$ \\
			$\left. L({f_{{\theta _{decoder}}}}({f_{{\theta _{encoder}}}}(px_i^*) + px_i^*),\overline {pt}) \right)$
			\ENDFOR
			\STATE $\delta^*$ = ${f_{{\theta _{encoder}}}}(px_{index({L_{\max }})}^*)$
			\RETURN $\delta^*$
		\end{algorithmic}
		
	\end{algorithm}
	
	\begin{algorithm}[tb]
		\caption{Dynamic trigger generation}
		\label{al3}
		\textbf{Input}: ${f_{{\theta _{encoder}}}}$; ${x_{train}}$(${x_{train}}$ is the training data)\\
		\textbf{Output}: ${\delta ^*}$(Dynamic trigger);
		
		\begin{algorithmic}[1]
			\STATE ${\delta ^*}$ = ${f_{{\theta _{encoder}}}}({x_{train}})$
			\RETURN{$\delta^*$}
		\end{algorithmic}
	\end{algorithm}
	
	\textbf{Step 4: Inference.} To execute an attack on a test input $x_{\text{test}}$, the attacker amplifies a trigger to a certain scale and integrates it into the test dataset. The neural network then classifies $x_{\text{test}}$ as the target class. Notably, the concept of trigger amplification during testing, as previously discussed in \cite{r9,r11}, may enhance resistance to backdoor detection by minimizing noise injection.
	
	\begin{figure*}[tb]
		\centering
		\includegraphics[width=1.95\columnwidth]{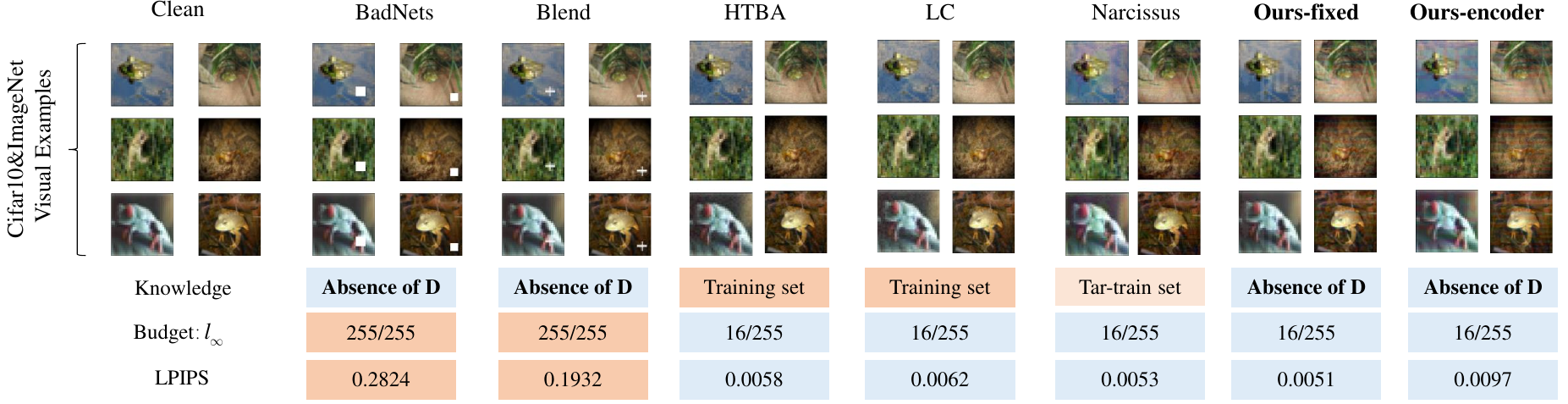} 
		\caption{Visual Comparison of Poisons}
		\label{fig_show_img}
	\end{figure*}
	
	\section{Experiments}
	
	\subsection{Experiments Settings}
	\textbf{General Setup.} The server system we used is Ubuntu 20.04, and it is equipped with two RTX 3090 GPUs as the hardware platform for conducting and evaluating experiments. In most experiments, we utilized the widely-adopted ResNet-18 architecture \cite{r5} for the target model. For the encoder, the U-Net architecture \cite{r28} was chosen, while ResNet-18 was used for the decoder's structure. To minimize the impact of network architecture on the outcomes, experiments were also conducted with different decoder architectures, such as VGG-16 \cite{r33} and VGG-19. Typically, the maximum poisoning ratio was set to 0.05\%. The $l_\infty$-norm bound for trigger generation was established at 8/255, and the poisoning level at 16/255, in line with standard practices in existing studies \cite{r9,r10}.

	\textbf{Datasets Setup.} Our attack was evaluated on three datasets: CIFAR-10 \cite{r29}, Tiny-ImageNet \cite{r30}, and TSRD \cite{r32}. Due to limitations in accessing training data, we selected POOD datasets corresponding to each victim model's dataset, specifically Tiny-ImageNet, CIFAR-10, and GTSRB \cite{r31}, ensuring no class overlap between the training and POOD datasets. Detailed information regarding the datasets used, the attack datasets, and the hyperparameters for each training pipeline is provided in Table \ref{tb3}.

	\textbf{Attack Setup.} We evaluated two variants of DFB (fixed trigger and encoder trigger) against existing clean-label attacks, namely HTBA and LC, implementing these methods as described in their respective original papers. \textbf{Note that this is an unfair comparison, as they can access the training set}. Additionally, we modified two standard tainted-label attacks, BadNets and Blend (referred to as BadNets-c and Blend-c, respectively), by contaminating only the target category while preserving its original label, thus adapting them for a clean-label context. \textbf{This is fair, as it is under the same conditions without access to the training set}. Furthermore, we conducted dirty-label attacks using BadNets and Blend, denoted as BadNets-d and Blend-d, for comparative analysis with our approach. \textbf{It should be noted, however, that comparing dirty-label attacks with clean-label methods is not entirely fair, as dirty-label attacks are more easily detectable by data validators}.

	\begin{table}[tb]
		
		\caption{Dataset, POOD dataset for attacks, and hyperparameter configuration}
		\label{tb3}
		\resizebox{0.47\textwidth}{!}{
			
			\begin{tabular}{c|c|c|c}
				
				Dataset & CIFAR-10 & Tiny-ImageNet & TSRD  \\  \hline
				Classes & 10            & 200                &  58    \\\hline
				
				Input Shape & 32*32   &64*64 & 64*64 \\ \hline
				
				POOD dataset for attacks & Tiny-ImageNet  & CIFAR-10  & GTSRB  \\ \hline
				
				Poison Ratio (\%)  & 0.05 (50/50000) & 0.025(25/100000)     & 0.4(20/4173)   \\ \hline
				
				Target Class & 3(Tree frog) & 6(frog)     & 2(Limit 30)   \\ \hline
				
				Epochs & 200 & 200 & 200\\ \hline
				Optimizer & SGD & SGD & SGD\\\hline
				Augmentation & (Crop, R-H-Flip) & (Crop, R-H-Flip) & (Crop, R-H-Flip)\\ 
				
		\end{tabular}}
		
	\end{table}

	\subsection{Evaluation}
	
	\textbf{Visual Comparison of Poisons.} In both CIFAR-10 and Tiny-ImageNet, We conducted a visual comparison with existing clean-label attacks, utilizing LPIPS for quantitative assessment. We found that in terms of image quality, our attack is comparable to existing clean-label attacks as shown in \ref{fig_show_img}

	\begin{figure}[tb]
		\centering
		\includegraphics[width=0.81\columnwidth]{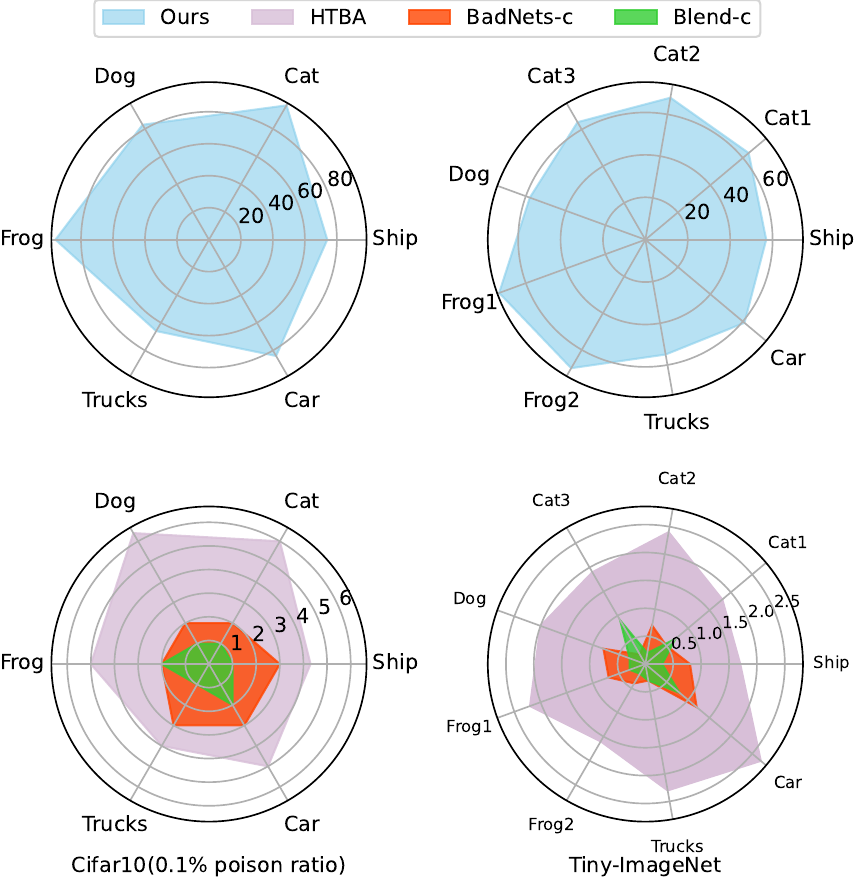} 
		\caption{Comparison of Attack Effects on Different Classes}
		\label{figatt}
	\end{figure}
	
	\textbf{Attack Effectiveness.} Our comparative results with other attack methods are shown in Table \ref{tb4}. On CIFAR-10, Tiny-ImageNet, and TSRD datasets, with poisoning rates of 0.1\%, 0.025\%, and 0.4\% respectively, DFB exhibits superior performance even under challenging conditions (i.e., without access to any training dataset). Utilizing fixed triggers from the POOD dataset, devoid of target-class features, and dynamically adjusting encoder triggers based on the training set, DFB achieves faster model injection. This results in a higher attack success rate (ASR) compared to LC and HTBA. In contrast to fair attacks like BadNets and Blend, our method shows higher success rates while preserving prediction accuracy on clean datasets. Further, in experiments attacking every class in CIFAR-10 and Tiny-ImageNet, given our restriction of no access to clean labels from the training set, our approach outperforms BadNet-c and Blend-c. We achieved the highest ASR on attacking all classes, where BadNet-c and Blend-c did not surpass a 10\% ASR. Additionally, our performance exceeded that of HTBA, as depicted in Figure \ref{figatt}.
	
	\textbf{Trigger Evaluation.} We conducted trigger evaluations on the CIFAR-10 and Tiny-ImageNet datasets. A clean ResNet network was trained, achieving test accuracies of 95.4\% on CIFAR-10 and 69.3\% on Tiny-ImageNet. We then applied various triggers to the entire training set and monitored their impact on CrossEntropyLoss, with results detailed in Table \ref{tbtrigger}. Our findings indicate that triggers from BadNets and Blend were less effective in modifying image features, posing challenges for neural network learning. In contrast, DFB-generated triggers, optimized using the POOD dataset, significantly obscured key features, leading to a marked increase in CrossEntropyLoss. This effectively validates the approach outlined in our proposed formula \ref{eq1}.

	\begin{figure}[tb]
		\centering
		\includegraphics[width=0.95\columnwidth]{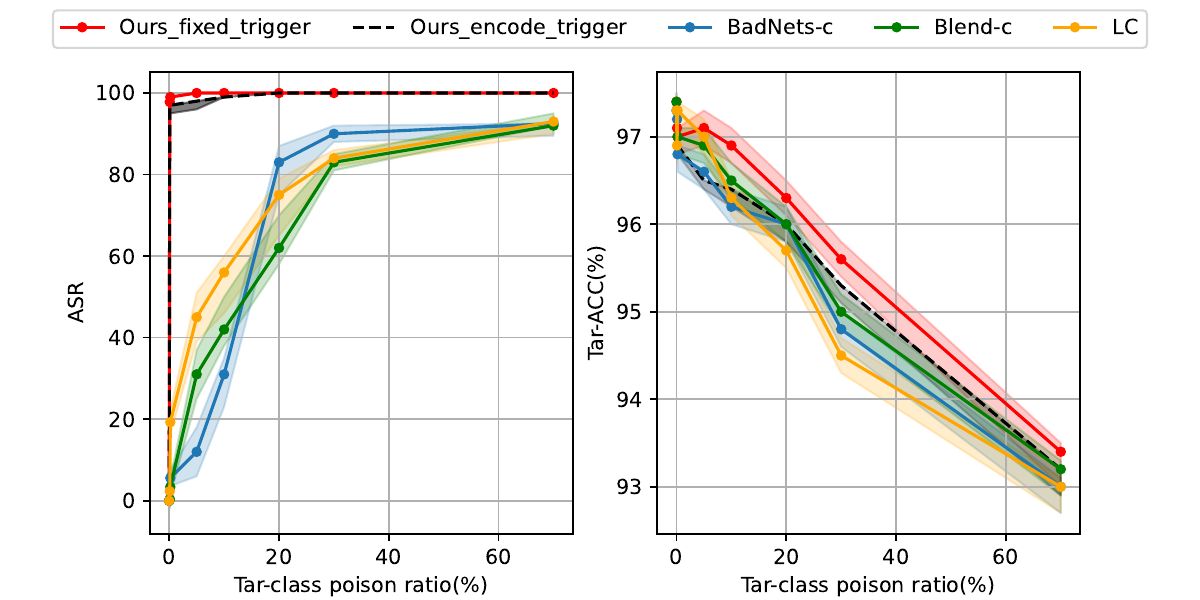} 
		\caption{Comparison of attack success rate and prediction accuracy of two DFB attack methods with other attack methods at different poisoning rates}
		\label{fig3}
	\end{figure}

	\textbf{Impact on Accuracy.} Figure \ref{fig3} illustrates the performance of two trigger generation methods for DFB, showing how prediction accuracy and attack success rate vary with increasing poisoning rates. We noted that as the poisoning rate escalated, DFB's attack success rate significantly outstripped those of LC, BadNets, and Blend. Additionally, DFB maintained comparable accuracy on clean datasets, akin to the performance of the "encode\_trigger" method. This suggests that a lower attack budget might have a less pronounced impact on the model's overall accuracy.
	
	\begin{table*}[t]
		
		\caption{Comparison of different triggers on CIFAR-10 and Tiny-ImageNet by observing their CrossEntropyLoss using clean classifiers (3 times average)}
		\label{tbtrigger}
		\centering
		\resizebox{1\textwidth}{!}{
			\begin{tabular}{c|c|c|c|c|c|c|c|c|c|c}
				\hline
				
				Dataset $ \rightarrow $& \multicolumn{5}{c|}{Cifar10} & \multicolumn{5}{c}{Tiny-ImageNet}\\ \hline
				\makecell[c]{Method $ \rightarrow $ \\ Class $\downarrow$} & Clean & BadNets & Blend & \textbf{Ours(Fixed)} & \textbf{Ours(Encoder)} & Clean & BadNets & Blend & \textbf{Ours(Fixed)} & \textbf{Ours(Encoder)} \\ \hline
				Frog &  0.0009& 0.0536& 0.0012& 0.8442 $\uparrow$ & 0.7884 $\uparrow$& 0.0058 & 0.0144& 0.0078& 5.2768$\uparrow$ & 7.1271$\uparrow$\\ 
				Cat & 0.0011 & 0.0621 & 0.0031& 2.207$\uparrow$ & 1.9873$\uparrow$& 0.0053 & 0.0087 &0.0083 & 5.5122$\uparrow$ & 6.3234$\uparrow$ \\ 
				Dog & 0.0009& 0.0058& 0.0022 & 2.136 $\uparrow$ &2.2322 $\uparrow$ & 0.0056 &0.0086& 0.0072 & 5.3471$\uparrow$ & 5.2036$\uparrow$ \\ 
				Truck &  0.0009& 0.0048 & 0.0026 & 0.6456 $\uparrow$ & 0.7673 $\uparrow$ & 0.0077&0.0111 & 0.0081 & 4.8921 $\uparrow$& 4.9737  $\uparrow$\\ 
				Ship & 0.0009&0.0109& 0.0254 & 0.7194$\uparrow$ & 0.6321$\uparrow$& 0.0018 & 0.0076& 0.0039 & 4.5813 $\uparrow$& 4.4727$\uparrow$ \\ 
				Car & 0.0011& 0.0065& 0.0041& 1.3612$\uparrow$ & 1.4351 $\uparrow$& 0.0051 & 0.0073& 0.0052&5.1392 $\uparrow$ & 5.0471$\uparrow$ \\ \hline 
				
		\end{tabular}}
	\end{table*}

\begin{figure}[tb]
		\centering
		\includegraphics[width=1\columnwidth]{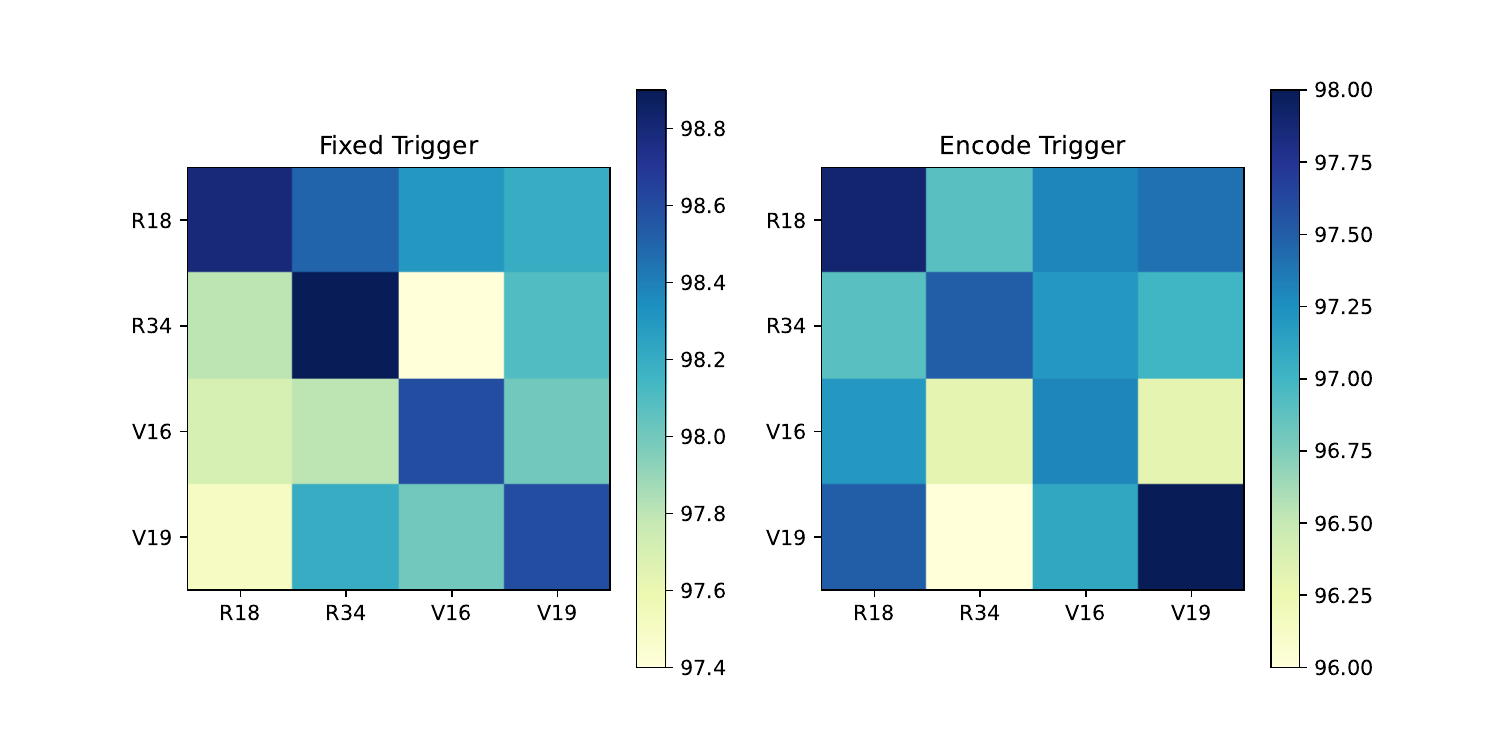} 
		\caption{The impact of ASR on different decoder network architectures}
		\label{fig4}
	\end{figure}

	\textbf{Impact of Different Network Architectures on Attacks.} In our study, we assessed the efficacy of two attack methodologies across a range of network architectures, modifying both the victim network and the decoder structure. We specifically employed ResNet18, ResNet34, VGG16, and VGG19 as the architectures for our decoder and victim networks. The findings, as presented in Figure \ref{fig4}, demonstrate that our attacks maintain their effectiveness irrespective of the varied architectural designs. We have observed that our generated triggers remain effective even when employing various neural network architectures as decoders. These triggers are also capable of effectively overlaying the features of the target class.

	\subsection{Ablation Study}
	
	To demonstrate the network's rapid learning capability for generating 'Without Target Features' in the POOD dataset, we conducted a series of ablation experiments.
	
\begin{figure}[tb]
		\centering
		\includegraphics[width=1\columnwidth]{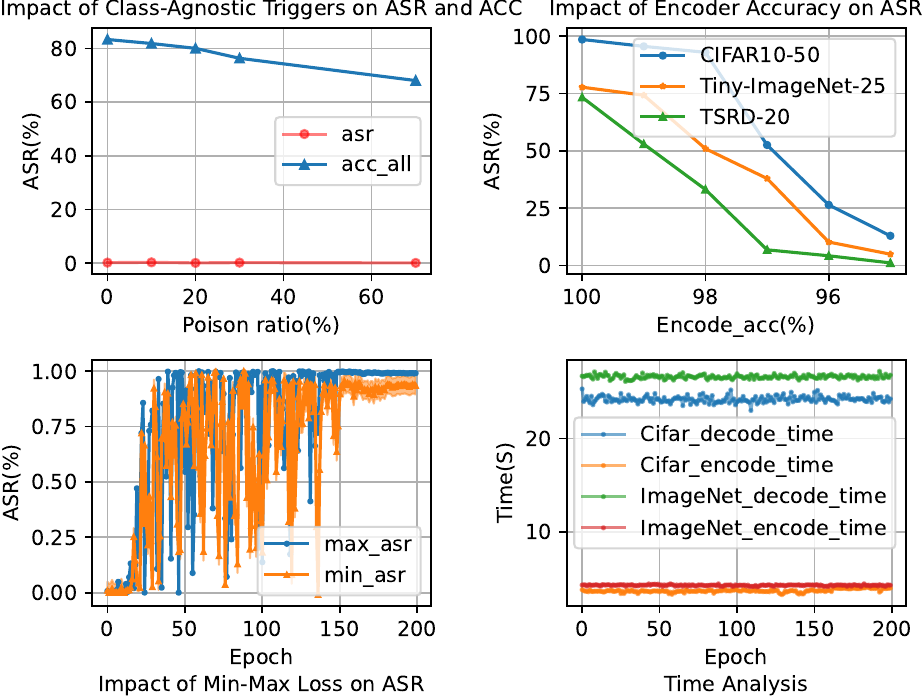} 
		\caption{Ablation study and time analysis}
		\label{fig5}
	\end{figure}

	Firstly, in 'Ablation Experiment 1: Impact of Class-Agnostic Triggers on ASR and ACC', we introduced triggers generated using Tiny-ImageNet into the TSRD dataset to observe their effects. This was intended to test our hypothesis: if correct, these triggers should be effective not only on CIFAR-10 but also on the TSRD dataset. Secondly, in 'Ablation Experiment 2: Impact of Encoder Accuracy on ASR', we evaluated the performance of triggers generated by the encoder, paying particular attention to their accuracy in the decoder. Lower accuracy in this context suggests that the triggers contain fewer features of the POOD target class. Lastly, in 'Ablation Experiment 3: Impact of Min-Max Loss on ASR', we experimented with triggers exhibiting maximum LOSS. As an extreme case, we also examined the impact of triggers with the minimum LOSS value. The results, depicted in Figure \ref{fig5}, indicated that the lower the accuracy of triggers generated by the encoder in the decoder, the lower the ASR. Furthermore, triggers generated using Tiny-ImageNet and injected into the TSRD dataset demonstrated no significant effect. In contrast, injecting triggers with the minimum LOSS value led to a decrease in ASR. Finally, our time analysis showed that the training duration is acceptable. Encoder training for one epoch takes only 3-4 seconds, while decoder training takes about 24 seconds.

	\subsection{Defense Resistance}
	
	We evaluated our defense against four typical backdoor defense methods: 1) Neural Cleanse, a defense method based on trigger synthesis; 2) the elimination of hidden backdoors in Deep Neural Networks (DNNs) through neuron pruning; 3) the detection of triggers using spectral features, as demonstrated by SentiNet; 4) STRIP, which detects potential backdoor triggers by utilizing perturbations and determines trigger regions through boundary analysis. In our experiments, we assessed the impact of our proposed 'DFB' attack on the effectiveness of these four representative defense solutions.
	
\begin{figure}[tb]
		\centering
		\includegraphics[width=0.7\columnwidth]{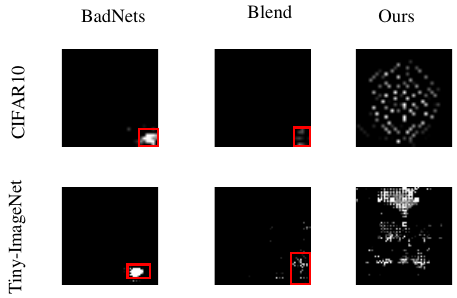} 
		\caption{The triggers reconstructed by Neural Cleanse.}
		\label{fig6}
	\end{figure}

\begin{figure}[tb]
		\centering
		\includegraphics[width=1\columnwidth]{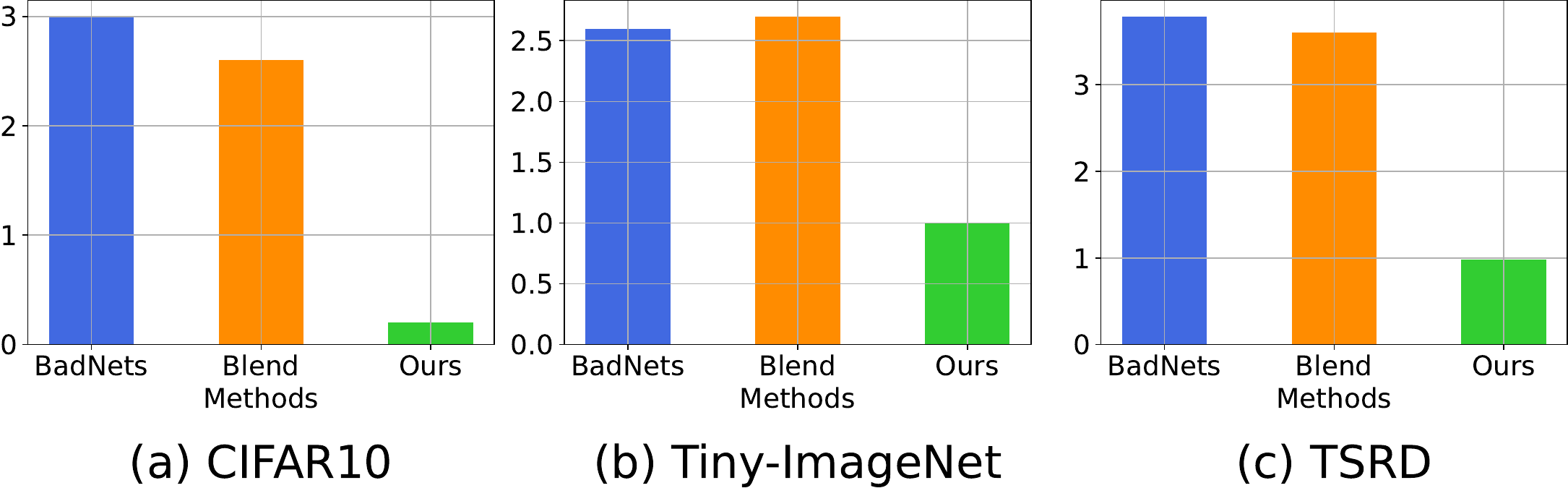} 
		\caption{The outliers computed by Neural Cleanse.}
		\label{fig7}
	\end{figure}

	\textbf{Resistance to Neural Cleanse.} The resistance of our method against Neural Cleanse is demonstrated in Figures \ref{fig6} and \ref{fig7}. Figure \ref{fig6} shows that our trigger, which spans the entire image, remains unrecoverable by Neural Cleanse. Furthermore, during the process of computing outliers, as shown in Figure \ref{fig7}, the values computed for our abnormal class are notably low. This indicates that our method exhibits strong resilience against Neural Cleanse.

\begin{figure}[tb]
		\centering
		\includegraphics[width=0.9\columnwidth]{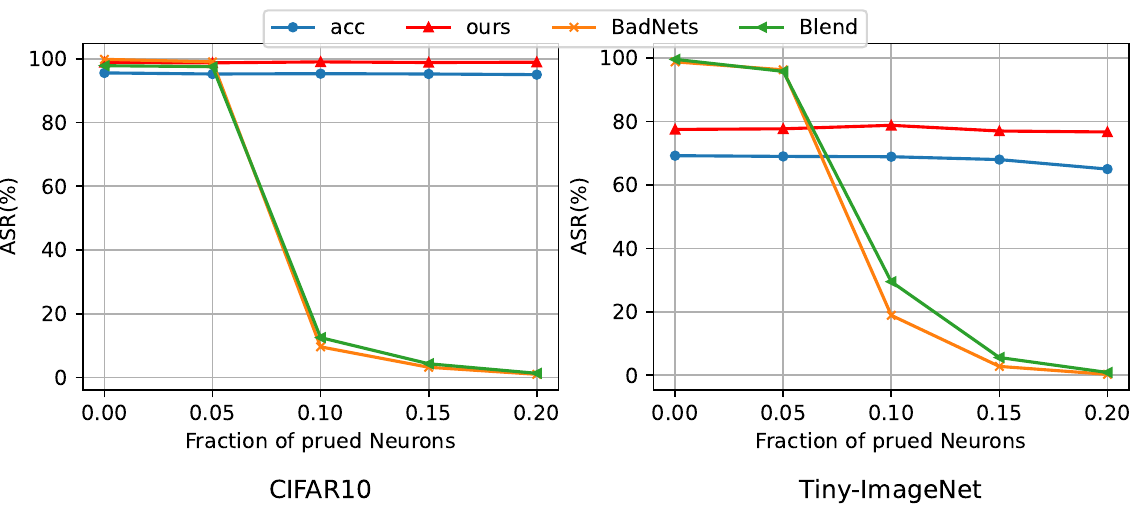} 
		\caption{Effect of Pruning Neurons at various rates on different attack methods}
		\label{fig8}
	\end{figure}

	\textbf{Resistance to Pruning.} We additionally tested the neuron pruning technique as a countermeasure against our attack. The results, as illustrated in Figure \ref{fig8}, show that our attack method demonstrates resilience against neuron pruning. A correlation between accuracy and the number of activated neurons is observable.
	
\begin{figure}[tb]
		\centering
		\includegraphics[width=1\columnwidth]{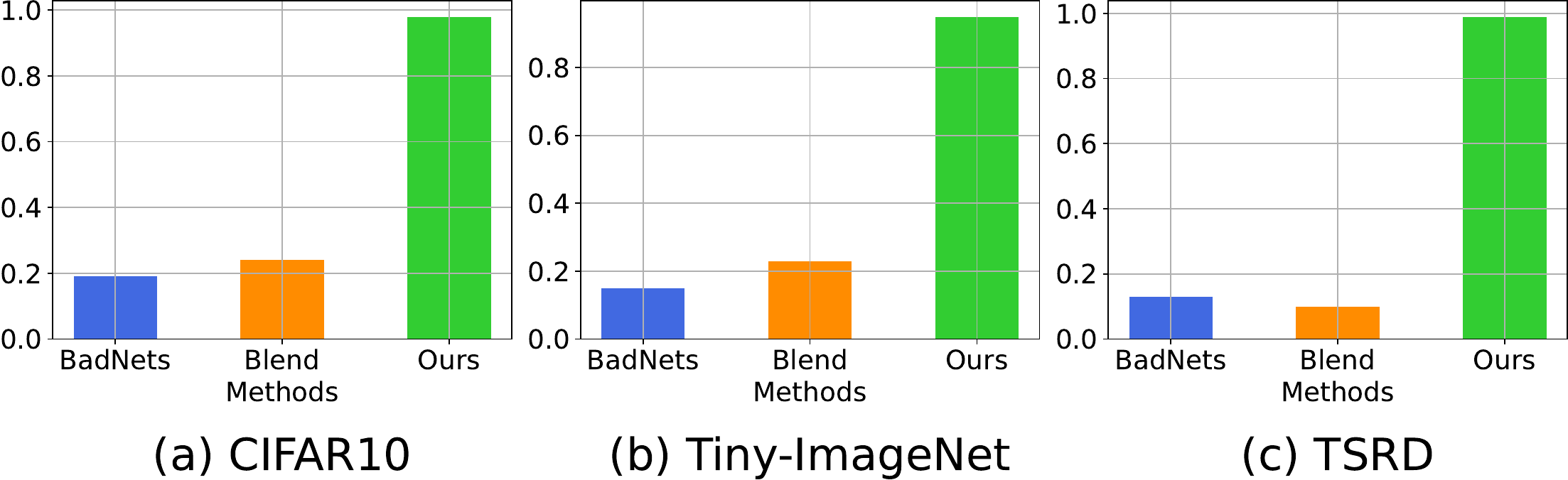}
		\caption{The entropy generated by poisoned images under the STRIP defense method}
		\label{fig9}
	\end{figure}

	\textbf{Resistance to STRIP.} We have evaluated our defense strategy against the STRIP method. During the training phase, the STRIP defense can analyze the dataset and overlay elements of the training set to detect abnormally low entropy, which is indicative of dataset contamination. This process is used to cleanse the dataset. Figure \ref{fig9} illustrates the effectiveness of our defense against STRIP. Our approach consistently outperforms others; for instance, methods like BadNets and Blend are effectively neutralized. Due to the minimal noise introduced by our triggering mechanism, our dataset maintains relatively low entropy, even when overlaid by STRIP.

\begin{figure}[tb]
		\centering
		\includegraphics[width=0.75\columnwidth]{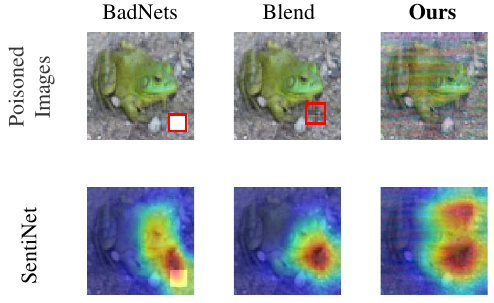} 
		\caption{Heatmaps generated by SentiNet based on the output of the backdoor model for poisoned images}
		\label{fig10}
	\end{figure}

	\textbf{Resistance to SentiNet.} SentiNet uses Grad-CAM to generate heatmaps reflecting the neural network's response to images, utilizing these heatmaps to ascertain if the network's labeling relies semantically on the images. Our approach was compared with BadNets and Blend, as illustrated in Figure \ref{fig10}. In the heatmaps of the backdoored networks' outputs, the triggers for BadNets and Blend are easily identified by the neural network. Conversely, the triggers generated by our method are more focused on semantic information, making them less detectable. This is further supported by our findings in the pruning defense: pruning neurons does not effectively eliminate our injected backdoor. The success rate of our backdoor attack is intricately linked to the accuracy of the network's predictions.

	\section{Conclusion}
	
	In this paper, we addressed the limitation of existing clean-label attacks, which typically rely on knowledge of either all or part of a category's training data. To overcome this, we introduced DFB, a novel clean-label backdoor attack that operates without requiring access to any training dataset. This method only necessitates knowledge of the clean labels for the categories under attack. Through extensive experimentation, we demonstrated the effectiveness of our approach. Additionally, our findings highlight that triggers, which are typically challenging for models to learn, can be effectively associated with target classes. Moving forward, our research agenda will focus on exploring and developing defense mechanisms against this type of sophisticated attack.

	\bibliographystyle{named}
	\bibliography{ijcai24}
	
\end{document}